\documentclass{svjour3}  
\smartqed  

\usepackage{amsmath, amssymb}
\usepackage{graphicx}
\usepackage{color}
\usepackage{graphicx}
\usepackage{verbatim}

\usepackage[colorlinks=true,linkcolor=blue,citecolor=blue]{hyperref}
\usepackage{braket}	
\newcommand{\be}{\begin{equation}}
\newcommand{\ee}{\end{equation}}
\newcommand{\bea}{\begin{eqnarray}}
\newcommand{\eea}{\end{eqnarray}}

\begin{document}

\title{Exact calculation of the large deviation function for $k$-nary coalescence}
\titlerunning{$k-$nary coalescence}
\author{R. Rajesh \and V. Subashri   \and Oleg Zaboronski}

\institute{V.Subashri \at
              	The Institute of Mathematical Sciences, CIT Campus, Taramani, Chennai 600113, India \\
              	Homi Bhabha National Institute, Training School Complex, Anushakti Nagar, Mumbai 400094, India\\
               	\email{subashriv@imsc.res.in}
          \and
          	R. Rajesh \at
		The Institute of Mathematical Sciences, CIT Campus, Taramani, Chennai 600113, India \\
              	Homi Bhabha National Institute, Training School Complex, Anushakti Nagar, Mumbai 400094, India\\
               	\email{rrajesh@imsc.res.in}
               	\and
          	Oleg Zaboronski\at
		Mathematics Institute, University of Warwick, Gibbet Hill Road, Coventry, CV4 7AL, UK\\
               	\email{olegz@maths.warwick.ac.uk}
}

\date{Received: \today / Accepted: }

\maketitle

\begin{abstract}
We study probabilities of rare events in the general coalescence process, $kA\rightarrow \ell A$, where $k>\ell$. For arbitrary $k, \ell$, by rewriting these probabilities in terms of an effective action, we derive the large deviation function describing the probability of finding $N$ particles at time $t$, when starting with $M$ particles initially. Additionally, the most probable trajectory corresponding to a fixed rare event is derived.
\end{abstract}

\section{\label{sec1-Introduction}Introduction}
Physical phenomena in which aggregation or coalescence of constituents is a dominant dynamical process are ubiquitous in nature. Examples include cloud formation~\cite{falkovich2002acceleration}, aerosol dynamics~\cite{williams1988unified,hidy1970dynamics,drake1972topics}, blood coagulation~\cite{guria2009mathematical}, dynamics of Saturn rings~\cite{brilliantov2015size,connaughton2018stationary,connaughton2017universality}, neurodegenerative disorders such as Alzheimer's disease~\cite{bekris2010genetics}, dynamics of polyelectrolytes~\cite{tom2016aggregation,tom2017aggregation}, ductile  fracture~\cite{pineau2016failure}, etc. A model that isolates the effect of aggregation is the cluster-cluster aggregation model (CCA) in which the only dynamics is aggregation of clusters of particles to form larger clusters.

CCA has been studied using different approaches. Historically, it has been studied using the Smoluchowski equation,  a deterministic mean-field, integro-differential equation for the rate of change of number of clusters of a particular size or mass (see Refs.~\cite{leyvraz2003scaling,aldous1999deterministic,krapivsky2010kinetic,wattis2006introduction} for reviews). The information of the physical system being modeled such as shape of clusters~\cite{puthalath2023lattice} as well as the transport properties of the constituents are incorporated into a collision kernel that describes the rate of collision between clusters of different sizes. The Smoluchowski equation ignores fluctuations, both spatial and stochastic. In lower dimensions, when spatial density fluctuations become dominant, the mass distribution in CCA has been studied using both analytical and numerical techniques~\cite{spouge1988exact,kang1984fluctuation,krishnamurthy2002kang,krishnamurthy2003persistence}.  However, these approaches are limited to analyzing the mean or typical mass distribution and its statistical low moments. They do not provide information about the probabilities of rare or atypical events, nor do they explain the pathways that lead to such events.

Recently, we developed a formalism  to calculate the large deviation function (LDF), which describes the probabilities for rare events, in CCA~\cite{rajesh2024exact}. This calculation was based on the  Doi-Peliti-Zeldovich-Ochinnikov (DPZO) method~\cite{doi1976second,doi1976stochastic,peliti1985path,ovchinnikov1978role,tauber2014,cardy2006reaction},  a path integral approach that is based on writing the probabilities in terms of an effective action. An exact expression of the LDF of CCA was obtained for three standard collision kernels, the constant (rate is independent of mass), sum (rate is sum of masses) and product (rate is product of masses) kernels. The LDF has a singularity for the product kernel which is indicative of the sol-gel transition, wherein the Smoluchowski equation no longer conserves mass beyond a certain time.  We could also determine the optimal evolution trajectories for a given rare event as solutions to the Euler-Lagrange equations that minimize the effective action~\cite{rajesh2024exact}. Other studies of the LDF in CCA include the study of the gelation transition in the product kernel  using large deviation theory in the probability literature (see~\cite{andreis2019large} and references within), and a Monte Carlo algorithm for numerically determining the LDF for arbitrary collision kernels~\cite{dandekar2023monte}. These results are for the case when the collisions were binary. In this paper we extend these results to $k$-nary collisions where $k$ particles aggregate in one event to form $\ell$ clusters, in particular the reaction $kA \xrightarrow{\lambda}\ell A$. We ignore the size of clusters, thus effectively studying the constant kernel problem.

We now briefly summarize what is known for the typical properties of the reaction $kA \xrightarrow{\lambda}\ell A$. Although the probability of more than two particles coalescing ($k>2$ is significantly smaller than that of two particles coalescing ($k=2$), higher order collisions may become important in certain cases, for example, when a structure formed by three particles exhibits more stability. Ternary collisions as the primary collision mechanism can be observed in experiments that were motivated by drug delivery~\cite{hinterbichler2015ternary}. Moreover, the reaction $kA \xrightarrow{\lambda}\ell A$  is one of the simplest examples of interacting particle systems that are far from equilibrium, making it a useful model for testing conceptual ideas. In the presence of diffusion, the upper critical dimension of the model is known to be $d_c=2/(k-1)$. Below $d_c$, the decay of mean density is dependent on the density fluctuations, and decays as $A_{k,\ell} t^{-d/2}$. Above $d_c$, the reaction is rate-limited and the mean density decays as $t^{-1/(k-1)}$. At $d_c$, mean density decays as $(t^{-1} \ln t)^{1/(k-1)}$. The prefactors for the power law decay can be computed as an $\epsilon$-expansion for $d<d_c$, and exactly at $d=d_c$~\cite{lee1994renormalization}. For $k=2$, using the empty interval method, the exact solution for the density can be found in one dimension~\cite{ben1998complete,ben1998inhomogeneous} and on the Bethe lattice~\cite{shapoval2018crossover}. Unlike density, the multi-particle correlations exhibit anomalous scaling in $d<d_c$. For $k=2$, this anomalous scaling can be determined using renormalization group methods in any dimension~\cite{munasinghe2006multiscaling} and rigorously in one dimension~\cite{munasinghe2006multi}. The scaling exponents are independent of $\ell$. This can be explicitly shown using field-theoretic methods~\cite{peliti1986renormalisation,lushnikov1987binary,lee1994renormalization,oshanin1995smoluchowski}. When mass is taken into account and for constant kernel, the exact result  may be found in one dimension~\cite{spouge1988exact} and using renormalisation group  in higher dimensions~\cite{krishnamurthy2002kang}. We also note that mass-dependent  $k$-nary aggregation has been studied, starting from the Smoluchowski equation in Refs.~\cite{jiang1990critical,jiang1994scaling,krapivsky2024gelation}.

In this paper, we are interested in rare events in $k-$nary coalescence, \textit{i.e.,} those events that occur at the tails of a probability distribution. The study of rare events is important because they could have significant impact despite their low likelihood of occurrence. Some common examples of rare events are natural disasters such as earthquakes and floods~\cite{ben2020localization}, financial black swan events~\cite{morales2020covid19}, and epidemics~\cite{clemenccon2015computer}. The mathematical framework for the systematic study of rare events is provided by the large deviation theory~\cite{touchette2009large}. The central focus of large deviation theory is the large deviation principle, that is, the probabilities of rare events decrease exponentially fast. The large deviation function, or rate function, captures information about large fluctuations (deviations) from the most probable or typical states of a system. The rate function can also be interpreted as a non-equilibrium generalization of entropy or free energy. 
  
In this paper, we compute the LDF for $k-$nary coalescence. We also calculate the most probable trajectory for a given rare event.

The remainder of this paper is organized as follows. In Sec.~\ref{model}, the model  is defined. In Sec.~\ref{exact}, the exact expression for the probability of $N$ particles remaining at time $t$  is derived for arbitrary $k,\ell$. In Sec~\ref{LDF}, the existence of a large deviation principle is proved, and the rate function is obtained, starting from the master equation.  The optimal evolution trajectories for arbitrary $k,\ell$ are also obtained. In Sec.~\ref{sec:asymp}, the asymptotic behavior of the rate function is described. Finally, we end with a summary and discussion in Sec.~\ref{sec:summary}.


		\section{Model}\label{model}
		Consider a system of particles which evolves in time through the generalized coalescence process, 
		\begin{equation}
		kA \xrightarrow{\lambda}\ell A,~~~~\ell<k,\label{eq:knary}
		\end{equation}
		where $A$ denotes a particle. Equation~(\ref{eq:knary}) describes the aggregation of $k$ particles into $\ell$ particles at constant rate $\lambda$. In other words, we model the $k$-nary coalescence by a continuous time Markov chain on the state space $\mathbb{N}_0$ consisting of non-negative integers, defined by the transition  $N\rightarrow N-(k-\ell)$ with the exponential rate $\lambda  {N\choose k} $.

		Each collision reduces the number of particles, $N(t)$, by $(k-\ell)$. The final absorbing state of this process contains $\ell,\ell+1,\dots k-1$ particles, depending on the value of the initial number of particles, $M$.
		
		In this paper, we study $P(M,N,t)$, the probability that exactly  $N$ particles remain at time $t$, given that there are $M$ particles initially. 
		The number of collisions that have occurred, $C$, is related to $N$ as
		\begin{equation}
		C=\frac{M-N}{k-\ell}.\label{eq:C1}
		\end{equation}

\section{Exact solution}\label{exact}
	It is possible to obtain the exact expression for $P(M,N,t)$ as a sum of exponentials for arbitrary $k,\ell$~\cite{dandekar2023monte}. Such an exact answer is possible because the collision rate after $C$ collisions is known exactly, unlike the situation when masses are assigned to particles and the collision rates depend on the mass distribution.
	For a given $k,\ell$, after $i$ collisions, $M-(k-\ell)i$ particles remain, and the total rate of collision for the $i$-{th} collision, $\mathcal{R}_{i}$, is therefore given by  \begin{equation}
	\mathcal{R}_{i}=\lambda {M-(k-\ell)(i-1)\choose k}.\label{rates}\end{equation} Using the exponential time distribution for waiting times, $P(\Delta t_i)=\mathcal{R}_i e^{-\mathcal{R}_i\Delta t_i}$,
	\begin{align}
	\begin{aligned}
	&P(M,N,t)=\int_{0}^{\infty}d\Delta t_1\int_{0}^{\infty}d\Delta t_2...\int_{0}^{\infty}d\Delta t_{C+1}  ~~\mathcal{R}_1 e^{-\mathcal{R}_1 \Delta t_1} \\&\mathcal{R}_2e^{-\mathcal{R}_2 \Delta t_2}\dots\mathcal{R}_{C}e^{-\mathcal{R}_{C}\Delta t_{C}}e^{-\mathcal{R}_{C+1}\Delta t_{C+1}}\delta\left(\sum_{i=1}^{C+1}\Delta t_i-t\right),
	\end{aligned}
	\end{align}
	where $C$ is as in Eq.~(\ref{eq:C1}).
	The final waiting time $\Delta t_{C+1}$ denotes the waiting time during which no collision occurs. The $\delta$-function constrains the sum of waiting times to the total time $t$. The Laplace transform of $\widetilde{P}(M,N,s)$, defined as 
	\begin{equation}
	\widetilde{P}(M,N,s)=\int_{0}^{\infty}dt e^{-st}P(M,N,t),
	\end{equation}
	is then
	\begin{equation}
	\widetilde{P}(M,N,s)=\prod_{i=1}^{C}\frac{\mathcal{R}_i}{\mathcal{R}_i+s}\frac{1}{\mathcal{R}_{C+1}+s}.
	\end{equation}
	Doing the inverse Laplace transform, we obtain
	\begin{equation}	
	P(M,N,t)=\left(\prod_{k=1}^{C}\mathcal{R}_k\right)\sum_{i=1}^{C+1}e^{-\mathcal{R}_i t} \prod_{j\neq i,\\{j=1}}^{C+1}\frac{1}{\mathcal{R}_j-\mathcal
		{R}_i}\label{soln}.
	\end{equation}
Though the exact expression for $P(M,N,t)$ can be obtained, it is not straightforward, either to derive the scaling for $N$ and $t$ with $M$ in the large deviation limit, nor  to derive the large deviation function directly from Eq.~(\ref{soln}). Also, it is not possible to obtain the optimal trajectory for a given rare event.  Instead, we will derive the large deviation function using the action formalism in Sec.~\ref{LDF}, and use the numerical evaluation of Eq.~(\ref{soln}) as a check for our results. In the process, we will also derive the optimal paths for rare events.

		\section{Results\label{LDF}} 
		\subsection{Master equation and effective action}
		The time evolution of $P(M,N,t)$ is described by the master equation 
		\begin{equation}
		\frac{dP(M,N,t)}{dt}=\lambda\left[\binom{N+k-\ell}{k} P(M,N+k-\ell,t)-\binom{N}{k}P(M,N,t)\right]. \label{main}
		\end{equation}
The first term on the right hand side of Eq.~(\ref{main}) is a gain term, which describes the creation of a state with $N$ particles, due to the aggregation of $k$ particles from a state with $N+k-\ell$ particles. The second term is a loss term, which describes the aggregation of $k$ particles from a state with $N$ particles. We now rewrite the calculation of $P(M,N,t)$ in terms of an effective action using the DPZO procedure~\cite{rajesh2024exact,greenman2018doi,peliti1985path}. Let
\begin{equation}
\ket{\psi(t)}=\sum_{N'=0}^M P(M,N^{\prime},t)\ket{N^{\prime}},\label{psi}
\end{equation} 
where $\ket{N}$ is the state with $N$ particles which is acted upon by creation and annihilation operators $a$ and $a^{\dagger}$,
and the number-of-particles operator $\hat{N}:=a^{\dagger} a$ as follows:
		\begin{align}
		&a\ket{N}=N\ket{N-1},\\
		&a^{\dagger}\ket{N}=\ket{N+1},\\
		&\hat{N}\ket{N}=N\ket{N},\\
		&[a,a^{\dagger}]=1.
		\end{align}
It is also useful to notice that $\braket{N^\prime|N}=N! \delta_{N',N}$.
		In terms of $\ket{\psi(t)}$, the master equation Eq.~(\ref{main}) can be rewritten as 
		\begin{equation}
		\frac{d\ket{\psi(t)}}{dt}=-\widehat{H}(a,a^{\dagger})\ket{\psi(t)}\label{master},
		\end{equation}
		where 
		\begin{equation}
		\widehat{H}(a,a^{\dagger})=-\frac{\lambda}{k!}(a^{\dagger \ell}-a^{\dagger k})a^k.
		\end{equation}
		Equation~(\ref{master}) has the solution $\ket{\psi(t_f)}=e^{-\widehat{H}t_f}\ket{\psi(0)}$, where $\ket{\psi(0)}=\ket{M}$ and $t_f$ is the final time. Multiplying Eq.~(\ref{psi}) on the left with $\bra{N}$, it is easy to see that
		\begin{equation}
		P(M,N,t)=\frac{\braket{N|\psi(t)}}{N!}.\label{master1}
		\end{equation}
	Briefly, to find a path integral representation of Eq.~(\ref{master1}), the evolution operator is first represented as an infinite
product using Trotter's formula, 
$$e^{-\widehat{H}(a^{\dagger},a)t}=\lim_{n\rightarrow \infty} \left(1-\widehat{H}(a^{\dagger},a)\frac{t}{n}\right)^n.$$  
Next the partition of the identity operator $\textbf{I}$ in terms of the complete set of eigenfunctions of the annihilation operator $a$ is inserted
between every pair of factors of $\left(1-\widehat{H}(a^{\dagger},a)\frac{t}{n}\right)$ entering the Trotter formula,
\begin{eqnarray*} 
e^{-\widehat{H}(a^{\dagger},a)t}=\lim_{n\rightarrow \infty} \left(1-\widehat{H}(a^{\dagger},a)\frac{t}{n}\right)
\textbf{I}\ldots \textbf{I} \left(1-\widehat{H}(a^{\dagger},a)\frac{t}{n}\right),\\
\mbox{where }
\textbf{I}=\int_{\mathbb{C}} \frac{dzd\tilde{z}}{\pi}e^{-|z|^2}\ket{z}\bra{z},~\ket{z}=e^{z a^{\dagger}}\ket{0}.
\end{eqnarray*}
Finally, the matrix elements $\bra{z^\prime} \left(1-\widehat{H}(a^{\dagger},a)\frac{t}{n}\right) \ket{z}$ are calculated
using the representation of the operators $a,a^{\dagger}$ in the basis of $(\ket{z})_{z\in \mathbb{Z}}$ (the so-called  holomophic representation):
$$
\ket{z}=e^{zw},~a^\dagger=w\cdot,~ a=\frac{\partial}{\partial w}, \braket{z | z^\prime}=e^{\overline{z}z^\prime},
 ~w,z,z^\prime\in \mathbb{C}
$$
where the inner product is defined for a pair of holomorphic functions $f,g$ as $\braket{f | g}=\int_{\mathbb{C}}\frac{dwd\bar{w}}{\pi}e^{-w\overline{w}} \overline{f(w)} g(w)$.
The final answer is
	\begin{equation}
	P(M,N,t)= \int \mathcal{D}\tilde{z}(t)\mathcal{D}z(t) e^{-S(z,\tilde{z},{t})},\label{eq:19}
	\end{equation}
	where the action is given by
	\begin{equation}
	S(z,\tilde{z},{t})=\int_0^{t} dt \left[\tilde{z}\dot{z}+\lambda H(z,\tilde{z})-N\ln z(t_f)\delta(t_f-t)-M\ln\tilde{z}(0)\delta(t)\right]+M+\ln N!,
	\end{equation}
	and the effective Hamiltonian $H$ is
	\begin{equation}
	H(z,\tilde{z})=-\frac{\big(\tilde{z}^\ell-\tilde{z}^k\big)z^k}{k!}. \label{eq:energy}
	\end{equation}
	
	\subsection{Existence of a large deviation principle}
	Defining $z\to zM^{\alpha}$, $\tilde{z}\to \tilde{z}M^{\beta}$ and $\tau= \lambda tM^{\gamma}$, and substituting in the action, we find that the choice $\alpha=1$, $\beta=0$ and $\gamma=k-1$ keeps the form of the action unchanged. The scaled action is then
	\begin{align}
	S(z,\tilde{z},{\color{red}\tau})=M\left[\int_0^{\tau} d\tau^{\prime} \Big[\tilde{z}\dot{z}+E(z,\tilde{z})\Big]-\phi\ln z(\tau)-\ln\tilde{z}(0)+z(0)\tilde{z}(0)+\phi\ln \frac{\phi}{e}\right],\label{master2}
	\end{align}
	where $\phi=N/M$. 
In the limit $M,N\to\infty$, keeping $\phi=N/M$ and $~\tau= \lambda M^{k-1}t$ fixed, since the action in Eq.~(\ref{master2}) is proportional to $M$, 
the functional integral Eq.~(\ref{eq:19}) is dominated by the minimum of the action. Thus, there exists a large deviation principle
	\begin{equation}
\lim_{M\to \infty}-\frac{\ln P(M,\phi M, \tau [\lambda M^{k-1}]^{-1})}{M}=f(\phi,\tau),
\end{equation}	 
where $f(\phi,\tau):=\min_{z,\tilde{z}}S(z,\tilde{z},\tau)$ is the rate function. We note that the number of particles formed after each collision, $\ell$, does not appear in the scaling of time. 
		\begin{figure*}
		\centering
	\includegraphics[width=1\linewidth]{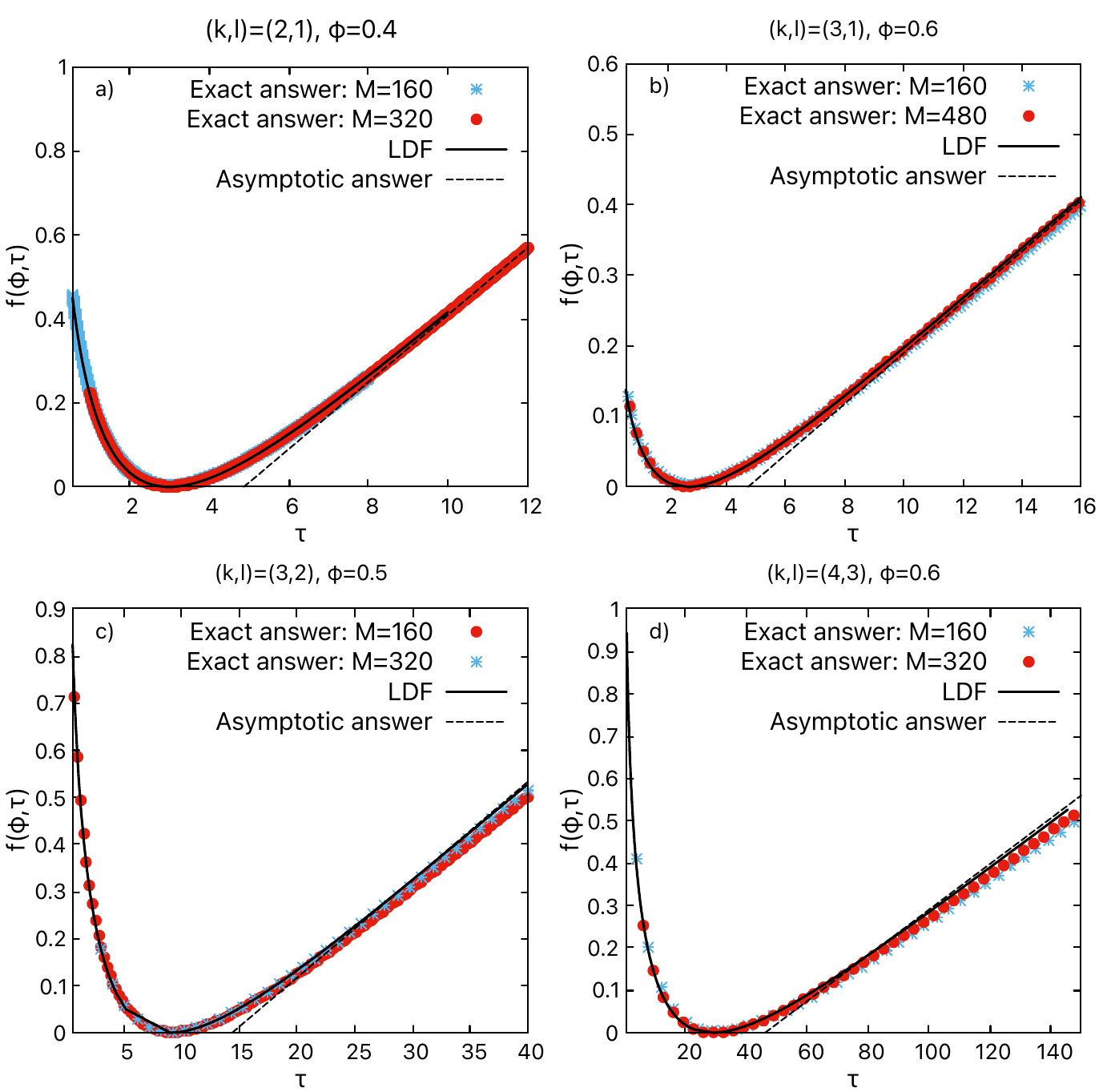}
	\caption{Large deviation functions $f(\phi,\tau)$ respect to $\tau$ are compared with the exact answer, Eq.~(\ref{soln}), for a) $(k,\ell)=(2,1), \phi=0.4$, b) $(k,\ell)=(3,1), \phi=0.6$, c) $(k,\ell)=(3,2), \phi=0.5$, d) $(k,\ell)=(4,3), \phi=0.6$. The agreement of the exact solution with the  large deviation function is better for larger values of the total mass $M$. The asymptotic answers for $\tau\to \infty$ are shown (broken black lines).}
	\label{action}
	\end{figure*}
	
			\begin{figure*}
		\centering
	\includegraphics[width=1\linewidth]{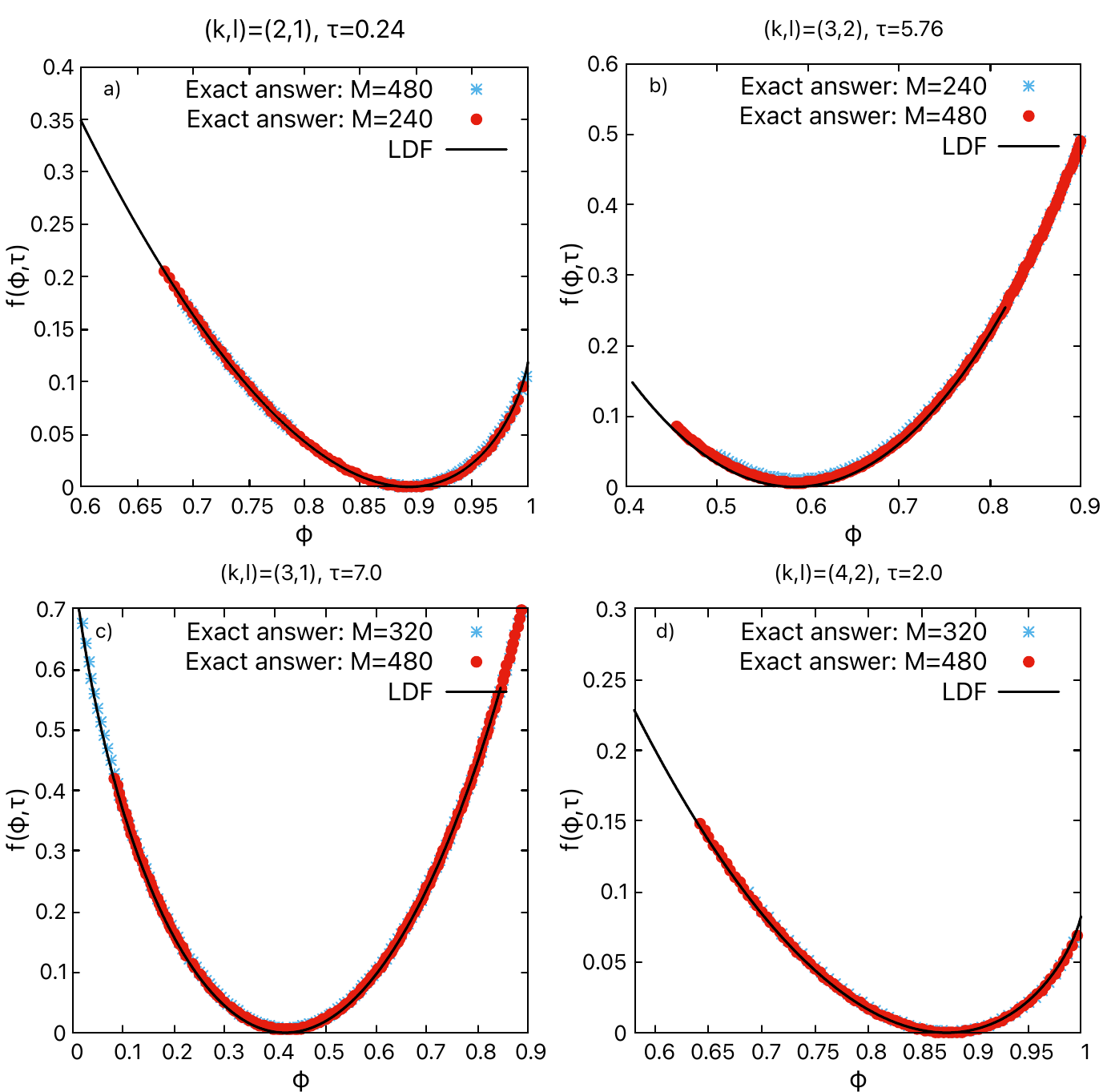}
	\caption{Large deviation functions $f(\phi,\tau)$ with respect to $\phi$ are compared with the exact answer Eq.~(\ref{soln}) plotted with respect to $\phi$ for a) $(k,\ell)=(2,1), \tau=0.24$, b) $(k,\ell)=(3,2), \tau=5.76$, c) $(k,\ell)=(3,1), \tau=7.0$, d) $(k,\ell)=(4,2), \tau=2.0$.}
	\label{action1}
	\end{figure*}

The Euler-Lagrange equations corresponding to the minimimum of the action, Eq.~(\ref{master2}) are given by	
	\begin{align}
	\dot{z}&=\frac{1}{k!}\Big(\ell\tilde{z}^{\ell-1}-k\tilde{z}^{k-1}\Big)z^k+\left(\frac{1}{\tilde{z}(0)}-z(0)\right)\delta(\tau),\label{z}\\ 
	\dot{\tilde{z}}&=-\frac{k}{k!} \Big(\tilde{z}^\ell-\tilde{z}^k\Big)z^{k-1}-\frac{\phi\delta(\tau-\tau_f)}{z(\tau_f)}=\frac{kE}{z}-\frac{\phi\delta(\tau-\tau_f)}{z(\tau_f)}.\label{zbar}
	\end{align}
	By integrating Eqs.~(\ref{z}) and (\ref{zbar}) about $\tau=0$ and $\tau=\tau_f$, we obtain the boundary conditions to be
	\begin{eqnarray}
	z(0)\tilde{z}(0)=n(0)=1,\label{bc1}\\
	z(\tau_f)\tilde{z}(\tau_f)=n(\tau_f)=\phi, \label{bc2}
	\end{eqnarray}
	where $n(\tau)=z(\tau)\tilde{z}(\tau)$ is the fraction of particles at time $\tau$.
	Note that the Euler-Lagrange equations conserve energy  $E:=H(z(\tau), \tilde{z}(\tau))$,  \textit{i.e.,}  $dE/d\tau=0$. Using 
Eq.~(\ref{eq:energy}), one can rewrite the energy  in terms of $n(\tau)$ to be 
	\begin{equation}
	E=\frac{n(\tau)^k-n(\tau)^\ell z(\tau)^{k-\ell}}{k!}.
\end{equation}
Knowing the values of the boundary conditions $n(0)$ and $n(\tau_f)$, Eqs.~(\ref{bc1}) and (\ref{bc2}) allows us to write $z(0)$ and $z(\tau_f)$, and consequently $\tilde{z}(0)$ and $\tilde{z}(\tau_f)$ in terms of $E$ and $\phi$. 
	 Additionally, integrating the first term in Eq.~(\ref{master2}) by parts, and using Eq.~(\ref{zbar}), the rate function for general $k,\ell$ is
	\begin{align}
	f(\phi,\tau)=-(k-1)E\tau-\!\!\frac{\phi}{k-\ell}\left[\ln\big(\phi^k-k!E\big)-\ell\ln\phi\right]+\frac{1}{k-\ell}\ln\big(1-k!E\big)+\phi\ln \phi,\label{eq:LDF}
	\end{align}	  
	where $E$ is a function of $\phi$ and $\tau$. It is determined by  the equation for the fraction of surviving particles $n$ 
which follows from the Euler-Lagrange equations (\ref{z}), (\ref{zbar}).  The corresponding  initial and final conditions follow from Eqs.~(\ref{bc1}), (\ref{bc2}). 
We will refer to the this equation
as the {\it instanton equation} and analyse it below.
	
	\subsection{The instanton equation}\label{sec:instanton}
	The instanton trajectory for $n(\tau)=z(\tau)\tilde{z}(\tau)$ that minimizes the action is is derived from Eqs.~(\ref{z}) and (\ref{zbar})  to be
	\begin{equation}
	\frac{dn}{d\tau}=(k-\ell)\Big[E-\frac{n^k}{k!}\Big],\label{instanton}
	\end{equation}
which needs to be solved subject to the initial conditon $n(0)=1$ and the final condition $n(\tau_f)=\phi$. As the equation is of the first order, the final condition yields an equation for the 'instanton energy' $E$. 
	Equation~(\ref{instanton}) implies that if $\tau\to (k-\ell)\tau$, then the instanton trajectory for fixed $k$ and different $\ell$ should be identical. 
\begin{figure*}
	\centering
		\includegraphics[width=0.8\linewidth]{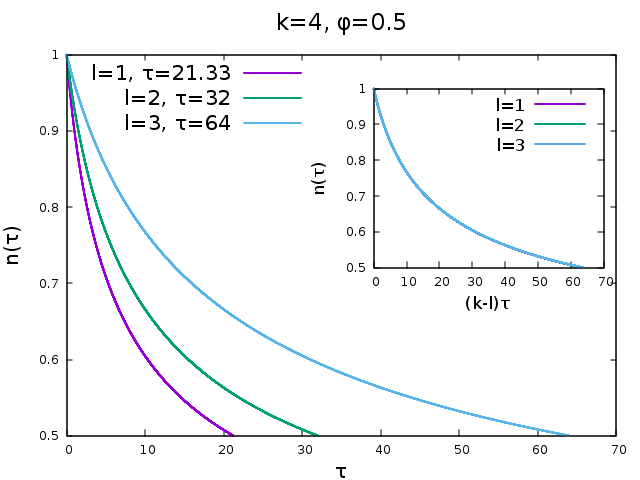}
		\caption{The instanton trajectories for  $k=3, \phi=0.3$ and $\ell=1,2$, plotted with respect to $(k-l)\tau$ collapse.}
		\label{eq:inst0}
	\end{figure*}
	When $E=0$, Eq.~({\ref{instanton}) for $n(\tau)$ reduces to the mean field equation for the mean number of particles. This corresponds to the mean field or the
'typical' solution which is followed by the system with the probability close to $1$:
	\begin{equation}
	n(\tau)=\Big[1+\frac{(k-\ell)(k-1)\tau}{k!}\Big]^{1/1-k},\label{typ}
	\end{equation}
It is seen that the typical trajectory achieves the fraction $\phi$ of the surviving particles at the time
\[
\tau_{typ}=\frac{k!}{(k-\ell)(k-1)}\left(\phi^{1-k}-1\right).
\]
It follows from (\ref{instanton}) that $E>0$ corresponds to rare events such that the time of reaching $\phi$ is smaller than $\tau_{typ}$,
$E<0$ - rare events reaching $\phi$ at the time larger than $\tau_{typ}$.
	The atypical trajectories corresponding to $E\neq 0$ can be obtained by solving Eq.~(\ref{instanton}). Rewriting it as
	\begin{equation}
	\frac{dx}{d\tau}=-\frac{(k-\ell)e_0^{k-1}}{k!}\Big[x^k-1\Big],\label{eq:inst4}
	\end{equation}
	where $x=n/e_0$ and $e_0^k=k!E$, factorising $1/(x^k-1)$ in terms of the $k$-th roots of unity, $(\omega_i)_{i=1}^k$, and then using the partial fraction decomposition, we obtain
	\begin{equation}
	\frac{1}{x^k(\tau)-1}=\frac{1}{\prod_{j=1}^k(x(\tau)-\omega_j)}=\sum_{m=1}^k \frac{A_m}{x(\tau)-\omega_m},\label{eq:inst1}
\end{equation}	 
where $\omega_j=e^{\frac{2\pi i j}{k}}$ and the coefficients $A_m$ are complex.
Solving for $A_m$, we obtain
\begin{equation}
A_m =\oint_{\Gamma_m} \frac{dz}{2\pi i }\frac{1}{z^k-1}=\frac{\omega_m}{k}\label{Am}.
\end{equation}
Here $\Gamma_m$ is a small contour around the $\omega_m$.

Substituting  Eq.~(\ref{Am}) into Eq.~(\ref{eq:inst1}),  one obtains the following implicit solution to Eq.~(\ref{eq:inst4}): 
\begin{equation}
\sum_{m=1}^k \omega_m \ln(x(\tau)-\omega_m)=-\frac{(k-\ell)(k!E)^{k-1}\tau}{(k-1)!}+c,
\end{equation}
where the constants $c$ and $E$ are fixed using the initial condition, $n(0)=1$ and the final condition $n(\tau_f)=\phi$. Solving for $c$, we obtain
\begin{equation}
	\sum_{i=1}^k\Omega_i\ln\frac{n(\tau)-\Omega_i}{1-\Omega_i}=-k(k-\ell)E\tau,\label{eq:inst3}
	\end{equation}
	where
	\begin{equation}
	\Omega_i=\omega_i (k!E)^{1/k}.
	\end{equation}
The instanton energy $E$ can be found by solving the equation
\begin{equation}\label{ieeq}
\sum_{i=1}^k\Omega_i\ln\frac{\phi-\Omega_i}{1-\Omega_i}=-k(k-\ell)E\tau_f,
\end{equation}
which can be analysed either numerically or analytically in the limit of $\tau_f>>\tau_{typ}$ or $\tau_f<<\tau_{typ}$
(see Sec.~\ref{sec:asymp} below).

It can be checked that for the constant kernel, $k=2$, $\ell=1$, we obtain
 $$n(\tau)=\sqrt{2E}\coth[ \sqrt{E/2}\tau+\tanh^{-1}\sqrt{2E}],$$ as derived in~\cite{rajesh2024exact}. The trajectories for various values of $(k,\ell)$ are shown in Fig.~\ref{inst2}.

		\begin{figure*}
	\centering
		\includegraphics[width=\linewidth]{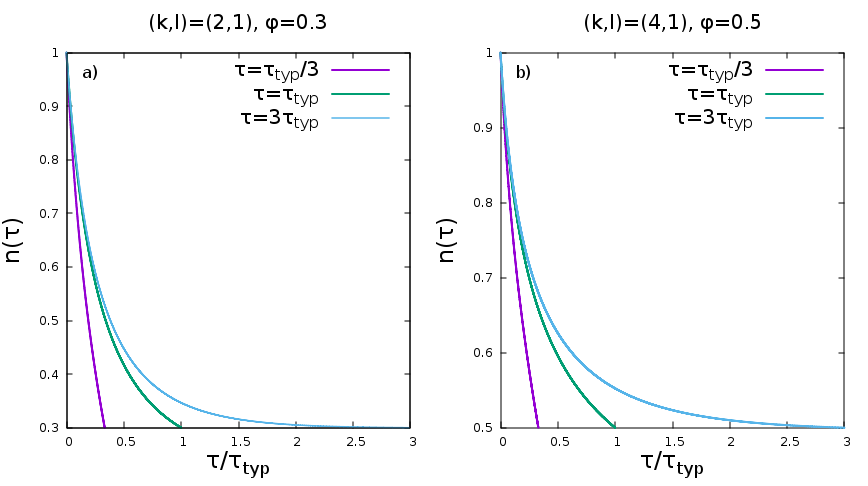}
		\caption{The instanton trajectories for a) Constant kernel, $(k,\ell)=(2,1), \phi=0.3$ and b) $(k,\ell=(3,2),\phi=0.7)$ are plotted for typical final time $\tau_{typ}$, which corresponds to $E=0$, as well as final times $\tau>\tau_{typ}$ and $\tau<\tau_{typ}$, which correspond to $E>0$ and $E<0$ respectively.}
		\label{inst2}
	\end{figure*}

\section{Comparison with exact answer}
In order to test the correctness of the rate function derived in Eq.~(\ref{eq:LDF}),  it is compared with the exact answer, Eq.~(\ref{soln}). The rate function is plotted  with respect to both $\tau$ and $\phi$, as shown in Figs~\ref{action} and \ref{action1} , and shows an excellent agreement with the exact answer. Figures~\ref{action}(a) and \ref{action1}(a) show the rate function for binary coalescence.

\section{Asymptotic analysis}	\label{sec:asymp}
It is possible to calculate exact asymptotics for  the large deviation in the limits  $\tau_f>> \tau_{typ}$ (anomalously slow evolution) and 
$\tau_f<<\tau_{typ}$ (anomalously fast evolution). Let us first consider the case $\tau_f>>\tau_{typ}$. 
In this limit, let us seek the solution to (\ref{ieeq}) in the form
\begin{equation}\label{elarge}
E=\frac{\phi^k}{k!}(1-\epsilon(\tau_f)),
\end{equation}
where $\epsilon(\tau_f)<<1$. Substituting this Ansatz into (\ref{ieeq}) one finds
\begin{equation}
\epsilon(\tau_f)=\exp\left(-\frac{k(k-\ell)\phi^{k-1}}{k!}\tau_f+O(1) \right).
\end{equation}
Substituting (\ref{elarge}) in Eq.~(\ref{eq:LDF}) and taking the limit $\tau_f\to\infty$, we obtain 
\begin{equation}
\lim_{\tau_f \to \infty} \frac{f(\phi,\tau_f)}{\tau_f} = \frac{\phi^k}{k!}.
\end{equation}
 Hence, the  asymptotic LDF for $k$-nary coalescence is the direct generalization of the constant kernel case~\cite{rajesh2024exact}.
			\begin{figure*}
		\centering
	\includegraphics[width=1\linewidth]{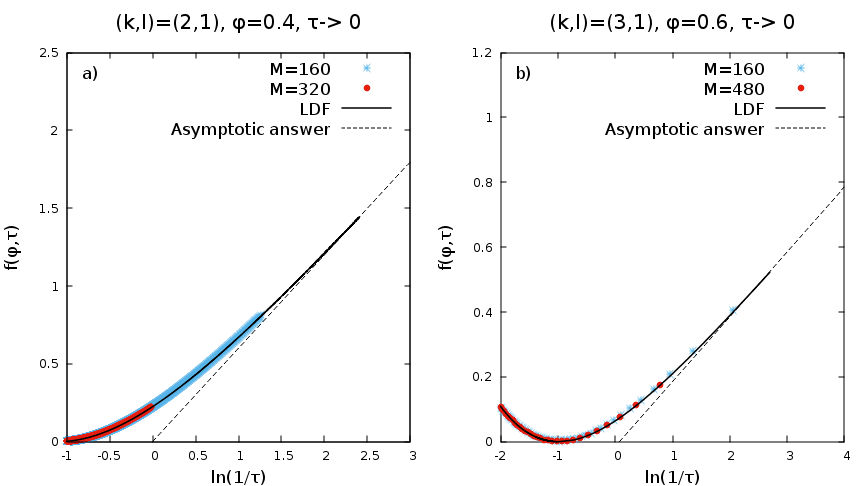}
	\caption{Large deviation functions $S(\phi,\tau)$ are plotted with respect to $\tau$ for a) $(k,\ell)=(2,1), \phi=0.4$ for $\tau\to\infty$, b) $(k,\ell)=(2,1), \phi=0.4$ for $\tau\to 0$, and with respect to $\ln(1/\tau)$ for c) $(k,\ell)=(3,1), \phi=0.6$, $\tau\to \infty$ and d)  $(k,\ell)=(3,1), \phi=0.6$, $\tau\to 0$.}
	\label{asymp}
	\end{figure*}
 
In the short time limit, $\tau<< \tau_{typ}$, the instanton energy $E$ is large and negative. Then $|\Omega_i|>>1$
 and the equation (\ref{ieeq}) takes the form
 \[
 k(1-\phi)+O(\Omega^{-1})=-k(k-\ell)E\tau_f.
 \]
 The solution is
 \begin{equation}
 E=-\frac{1-\phi}{(k-\ell)\tau_f}(1+O(\tau_f^{1/k})).
\end{equation}  
Substituting this solution in the LDF (\ref{eq:LDF}) and simplifying, we obtain
\begin{align}
\lim_{\tau_f \to 0} \frac{f(\phi,\tau_f)}{\ln (\tau_{typ}/\tau_f)} = \frac{1-\phi}{k-\ell}.\label{short}
\end{align}	

Figures~\ref{action} and \ref{asymp} show that the asymptotic answers for $\tau_f>>\tau_{typ}$ and 
$\tau_f<< \tau_{typ}$ respectively, are in excellent agreement with the exact answers. 

\section{Summary and discussion \label{sec:summary}}

To summarize, we derived the large deviation function for the general coalescence process, $k A \to \ell A$, for arbitrary $k >\ell$  using the path integral approach. The solution minimizing the action allowed us to determine the optimal trajectory for each rare event.

For the reaction  $k A \to \ell A$, it is possible to write an exact expression for $P(M,N,t)$ as a sum of exponentials. However, it is not straightforward to derive the large deviation function from this expression, neither  are the scaling variables obvious. Using the Doi-Peliti-Zeldovich method helps us to circumvent these issues. First the scaling variables become obvious, second we are able to determine the exact expression for the large deviation function and third,  the optimal trajectories for rare and typical events can be obtained.

The formalism used in this paper is generalizable to reaction diffusion systems in  higher dimensions, where a term related to transport of the clusters, such as a diffusion term would appear in the Euler Lagrange equations. Solvability remains as issue and a promising area for future research.  The formalism can also be generalized to coalescence with input~\cite{alemany1995inter,connaughton2004stationary,connaughton2005breakdown,connaughton2006cluster}, or with branching~\cite{ben1998diffusion,ben1998fisher,ball2012collective}, which could exhibit interesting features such as oscillations and steady states.


\end{document}